\documentclass[12pt]{article}
\usepackage{color}
\usepackage{bm}
\usepackage{slashed}
\usepackage{varioref}
\usepackage{xcolor} 
\usepackage{graphicx}
\usepackage{amsmath}
\usepackage{amssymb}
\usepackage{amsfonts}
\usepackage{amsthm}
\usepackage{array}
\usepackage{lineno}
\usepackage{multicol}
\usepackage{amsmath}
\usepackage{tensor}
\usepackage{mathtools}

\usepackage[
      colorlinks=true,
      linkcolor=blue,
      urlcolor=blue,
      filecolor=blue,
      citecolor=red,
      pdfstartview=FitV,
      pdftitle={},
      pdfauthor={},
      pdfsubject={},
      pdfkeywords={},
      pdfpagemode=None,
      bookmarksopen=true
]{hyperref}

\usepackage{epsfig}

\usepackage{hyperref}
\usepackage{float}
\usepackage{graphicx}
\usepackage{caption}
\usepackage{subcaption}

\def\beq{\begin{eqnarray}}
\def\eeq{\end{eqnarray}}

\def\r{\rho}

\def\be{\begin{equation}}
\def\ee{\end{equation}}
\def\bea{\begin{eqnarray}}
\def\eea{\end{eqnarray}}

\def\be{\begin{equation}}
\def\ee{\end{equation}}
\def\bea{\begin{eqnarray}}
\def\eea{\end{eqnarray}}

\newcommand{\rom}[1]{\mathrm{#1}}

\def\nn{\nonumber}

\numberwithin{equation}{section}

\begin{document}

\begin{center}
{\large \bf Tidal Forces in Kerr-AdS and Grey Galaxies}

\vspace{1cm}
Anand Balivada$^1$, Pius Ranjan Padhi$^{1,2}$,  and Amitabh Virmani$^1$

\vspace{1cm}

$^{1}$Chennai Mathematical Institute (CMI), \\ H1 SIPCOT IT Park, Kelambakkam, Tamil Nadu, India 603103\\
\vspace{0.5cm}
$^{2}$National Institute of Science Education and Research (NISER), Bhubaneswar, \\ P.O. Jatni, Khurda, Odisha, India 752050

\vspace{0.5cm}

\texttt{pius.padhi@niser.ac.in, anandb@cmi.ac.in, amitabh.virmani@gmail.com}
\end{center}

\begin{abstract}

In a recent paper [arXiv:2305.08922] \cite{Kim:2023sig}, it has been proposed that the endpoint of the Kerr-AdS superradiant instability  is a Grey Galaxy. The conjectured solutions are supposed to be made up of a black hole with critical angular velocity in the centre of AdS, surrounded by a large flat disk of thermal bulk gas that revolves around the black hole. In the analysis of the proposed solutions so far, gravitational effects due to the black hole on the thermal gas have been neglected.  A way to estimate these effects 
is via computing tidal forces. With this motivation,  we study tidal forces on objects moving in the Kerr-AdS spacetime.  To do so, we construct a parallel-transported orthonormal frame along an arbitrary timelike or null geodesic.  We then specialise to the class of fast rotating geodesics lying in the equatorial plane, 
and estimate tidal forces on the gas in the  Grey Galaxies, modelling it as a collection of particles moving on timelike geodesics.  
We show that the tidal forces are small (and remain small even in the large mass limit), thereby providing additional support to the idea that the gas is weakly interacting with the black hole.

\end{abstract}

\newpage

\tableofcontents

\newpage


\section{Introduction and motivation}

Black holes in anti-de Sitter space are of great importance in the gauge/gravity duality. Among other things, they determine the approach to equilibrium in the dual field theory. Detailed analysis \cite{Kunduri:2006qa, Cardoso:2013pza} of the perturbations of the Kerr-AdS black holes in four dimensions shows that unstable modes in Kerr-AdS are always associated to the superradiant instability and they appear only for black holes with $\Omega L > 1$, where $\Omega$ is the angular velocity of the black hole in the asymptotically static frame and $L$ is the AdS length. An important open question concerns the time evolution and the endpoint of the superradiant instability. Arguably, this is one of the most interesting open questions in the black hole perturbation physics, and has attracted a lot of attention in the last decade \cite{Dias:2015rxy, Niehoff:2015oga, Green:2015kur, Chesler:2018txn, Ishii:2021xmn, Chesler:2021ehz}.

In a recent paper \cite{Kim:2023sig}, it is conjectured that the endpoint of the superradiant instabilities 
for Kerr-AdS black holes in four dimensions
is a Grey Galaxy. Grey Galaxies are a (proposed) new class of coarse-grained solutions to Einstein's equations. The conjectured solutions are supposed to be made up of a black hole with critical angular velocity $\Omega L = 1$ at the centre of AdS, surrounded by a large flat disk of thermal bulk gas that revolves around the black hole. The key point is that the gas carries a finite fraction of the total energy and angular momentum.

Given various results in the literature, see, e.g., \cite{ Niehoff:2015oga, Hawking:1998kw}, the proposed picture is well motivated.  However, the picture is put forward by (essentially) analysing properties of the thermal bulk gas in the absence\footnote{The gas is far away from the black hole. This is how the proposal is justified.} of the black hole\footnote{The back reaction of the gas in \cite{Kim:2023sig} is computed only on a ``portion of global AdS'', neglecting the ``black hole tail''.}. More precisely, we do not know the nature of the gravitational forces acting on the gas in the Grey Galaxies. In the small Newton's constant expansion scheme of \cite{Kim:2023sig}\footnote{We wish to emphasise that the Grey Galaxies proposal is to be thought of in the large $N$ limit, where $N$ is the rank of the gauge group of the dual field theory via the AdS/CFT correspondence. In four-dimensions, the Newton's constant scales as $G_N \sim N^{-\frac{3}{2}}$ for AdS$_4$. With that in mind, in our paper $G_N$ is best thought of as a small parameter.}, these are small effects. However, for the construction of the full non-linear back-reacted solutions, this issue needs to be addressed. In this paper, we present a crude analysis that estimates gravitational effects on the gas in the Grey Galaxies. Throughout the paper we have in mind that the Newton's constant $G_N$ in small.

Often, the study of geodesics around black holes gives a first glance at the dynamical processes that have the black hole as an equilibrium. In most cases, the picture obtained is surprisingly precise. In this light, we propose that a crude but reasonable model for the gas in the Grey Galaxies is of timelike bound geodesics in the equatorial plane of the Kerr-AdS black hole. In this model gravitational forces acting on the gas---modulo various caveats about the classical vs quantum nature of the gas---can be estimated by studying tidal forces on relevant bound geodesics.

  With this motivation, we study tidal forces on a class of geodesics in Kerr-AdS spacetime. We show that the tidal forces are small for geodesics relevant for Grey Galaxies---in the sense made precise below---thereby providing additional support to the idea that the gas in the Grey Galaxies is weakly interacting with the black hole.

The question of computing tidal forces in Kerr-AdS is also of some intrinsic interest. In a general space-time, the equations of parallel transport are very difficult to solve explicitly even along known geodesics. However, in the particular case of  Kerr-AdS spacetime, the special separability properties that permit the integration of the geodesic equations can also be used to provide an explicit solution to the problem of constructing a parallel-transported orthonormal frame along an arbitrary time-like or null geodesic. This allows one to compute tidal forces on particles moving along these geodesics and also obtain the Penrose limit along null geodesics.

In  the literature \cite{Frolov:2017kze, Kubiznak:2008zs}, it is well appreciated that such calculations are doable, but explicit expressions have never been worked out except for the Kerr black hole. Motivated by  Grey Galaxies, we are led to make the formal results of this literature explicit for the case of 
Kerr-AdS spacetime. In this work, we follow the approach of Marck \cite{Marck1} who worked out the tidal tensor for arbitrary timelike \cite{Marck1} and null geodesics \cite{Marck2} for Kerr black hole and adapt his methodology to our problem. The earlier  formal results were obtained in a similar vein \cite{Kubiznak:2008zs}, with some slight technical variations. 
Among other applications, we note that the results of Marck \cite{Marck1, Marck2} have recently found applications in the study of tidal deformations of a binary system near a Kerr black hole \cite{Camilloni:2023rra}.

The rest of the paper is organised as follows. 
In section 
\ref{sec:geodesics}, we present a discussion on circular timelike and null orbits in the Kerr-AdS spacetime. Section
\ref{sec:main_results} consists of our main technical results, where we present a parallel-transported orthonormal frame for a general timelike or null geodesic. We also present expressions for tidal tensors along these geodesics.  In section 
\ref{sec:equatorial_geodesics}, we specialise to the equatorial plane. In section
\ref{sec:grey_galaxies}, we interpret our results in the context of Grey Galaxies. We end with a brief summary of our results in section
\ref{sec:conclusions}. While working on this project, we discovered a particularly simple presentation of deriving the inner most stable circular orbits (ISCOs) for the asymptotically flat Kerr black hole. To the best of our knowledge, such a presentation has not been been given before. Standard derivations tend to be algebraically cumbersome \cite{Bardeen:1972fi, Chandrasekhar:1985kt}. Ours is relatively less so.\footnote{See also ref.~\cite{Bardeen:1973tla}, where ``some simplified formulas derived by S.~Teukolsky" are given. Most likely, our derivation is the same as the one alluded to there.} We present this discussion in appendix
\ref{sec:Kerr_ISCO}.

\section{Kerr-AdS geodesics}
\label{sec:geodesics}
General timelike and null geodesics for Kerr-AdS were studied in a detailed paper 
 by Hackmann, Kagramanova,  Kunz, and L\"ammerzahl~\cite{Hackmann:2010zz}. Since circular timelike and null orbits in the equatorial plane are of crucial important for our study, we start with a presentation of these results. In five dimensions, a discussion of the type we are interested in was presented in \cite{Delsate:2015ina}.

 The Kerr-AdS metric in the Boyer-Lindquist like coordinate system takes the form \cite{Carter:1968ks, Chambers:1994ap}:
\begin{equation}
d s^2= -\frac{\Delta_r}{\chi^2 \r^{2}}\left(d t-a \sin ^2 \theta d \phi\right)^2+\frac{\r^{2}}{\Delta_r} d r^2 +\frac{\Delta_\theta \sin ^2 \theta}{\chi^2 \r^{2}}\left(a d t-\left(r^2+a^2\right) d \phi\right)^2+\frac{\r^{2}}{\Delta_\theta} d \theta^2 \label{metric}
\end{equation}
where,
\begin{align}
&\Delta_r  =\left(1-\frac{\Lambda}{3}r^2\right)\left(r^2+a^2\right)-2 m r, &    & \Delta_\theta  =1+\frac{a^2 \Lambda}{3} \cos ^2 \theta, \\
  &  \chi  =1+\frac{a^2 \Lambda}{3} = 1-\frac{a^2}{L^2} , &   & \r^{2}   =r^2+a^2 \cos ^2 \theta,
\end{align}
with $\Lambda = - \frac{3}{L^2}$. This metric asymptotically approaches global AdS. However, the asymptotic structure is not manifest in these coordinates. The coordinate frame $(t, r, \theta, \phi)$ rotates at infinity with angular velocity $\Omega_\infty = - \frac{a}{L^2 \chi}$. If we introduce,
\begin{align}
&T = \frac{t}{\chi}, &
&\Phi = \phi + \frac{a}{L^2}  \frac{t}{\chi}, \\
& R = \frac{\sqrt{L^2 (r^2 + a^2) - (L^2 + r^2) a^2 \cos ^2 \theta }}{L \sqrt{\chi}} , &
&\cos \Theta =\frac{L \sqrt{\chi} r \cos \theta}{\sqrt{L^2 (r^2 + a^2) - (L^2 + r^2) a^2 \cos ^2 \theta }},
\end{align}
then the Kerr-AdS metric approaches global AdS as $R \to \infty$,
\be
ds^2 = - \left( 1 + \frac{R^2}{L^2}\right) dT^2 +  \left(1 + \frac{R^2}{L^2} \right)^{-1} dR^2 + R^2 (d \Theta^2 + \sin^2 \Theta d \Phi^2).
\ee
The ADM mass and angular momentum of the black hole are related to the mass parameter $m$ and the rotation parameter $a$ via
\bea
& &M_\rom{ADM} = \frac{m}{G_N\chi^2}, \\
& & J_\rom{ADM} = \frac{m a}{G_N\chi^2},
\eea
where $G_N$ is the Newton's constant. 
We note that for fixed $m$, in the limit $a \to L$, both the ADM mass and angular momentum of the black hole become large.\footnote{We restrict ourselves to the parameters ranges $m> 0$ and $|a| < L$. For a more complete discussion of the parameter ranges we refer the reader to refs.~\cite{Kim:2023sig,Cardoso:2013pza}.}

We can identify four constants of motion for the geodesic motion. Two corresponding to the energy per unit mass $E$ and the $z-$component of the angular momentum $J$. We have\footnote{We define the Killing energy $E$ and the angular momentum $J$ with reference to the $(t, \phi)$ coordinates. It is most convenient to do so at this stage. Later when we discuss Grey Galaxies we will work with energy and angular momentum with reference to the $(T, \Phi)$ coordinates.}
\begin{align}
&  E :=  - g_{tt} \dot{t} - g_{t\phi} \dot \phi, \\
& J := g_{\phi \phi} \dot \phi + g_{t \phi} \dot {t}, 
\end{align}
where dots denote derivatives with respect to the proper time $\tau$ (or an affine parameter in the case of a null geodesic). The third constant of motion is given by the normalisation condition 
\be
\mu^2 = - g_{\mu \nu} \dot{x}^\mu \dot{x}^\nu, \label{mu2}
\ee
with $\mu^2 = 0$ for null geodesics and $\mu^2 =1$ for timelike geodesics. The fourth constant is the famous Carter's constant,
\be
K_{\mu \nu} \dot{x}^\mu \dot{x}^\nu = K,
\ee
where an expression for the Killing tensor $K_{\mu \nu}$ is given by the square of the Killing-Yano tensor, $K = f_{\mu \rho} f^{\rho}{}_{\nu}$. The Killing-Yano tensor is written below,  eq.~\eqref{KillingYano}. 
 We arrive at the separated equations in the following convenient form \cite{Frolov:2017kze, Hackmann:2010zz}:
\begin{align}
  &  \dot{t}=\frac{\chi^2}{\r^{2}}\left[(r^2+a^2)\frac{\{(r^2+a^2)E-aJ\}}{\Delta_r}-a\frac{(aE\sin^2\theta-J)}{\Delta_\theta}\right],
    \label{tgeo} \\
 &   \dot{\phi}=\frac{\chi^2}{\r^{2}}\left[\frac{a}{\Delta_r}\{(r^2+a^2)E-aJ\}-\frac{(aE \sin^2\theta-J)}{\Delta_\theta  \sin^2\theta}\right],
    \label{phigeo} \\
&    \r^4\dot{r}^2=\{(r^2+a^2)E-aJ\}^2\chi^2-\Delta_r(\mu^2r^2+K),
    \label{rgeo} \\
 &   \r^4\dot{\theta}^2=\Delta_\theta(K-\mu^2a^2\cos^2\theta)-\chi^2\left(aE\sin\theta-\frac{J}{\sin\theta}\right)^2.
    \label{thetageo}
\end{align}

 \subsection{Circular null orbits}

  \begin{figure}[t]
\begin{center}
  \includegraphics[width=10cm]{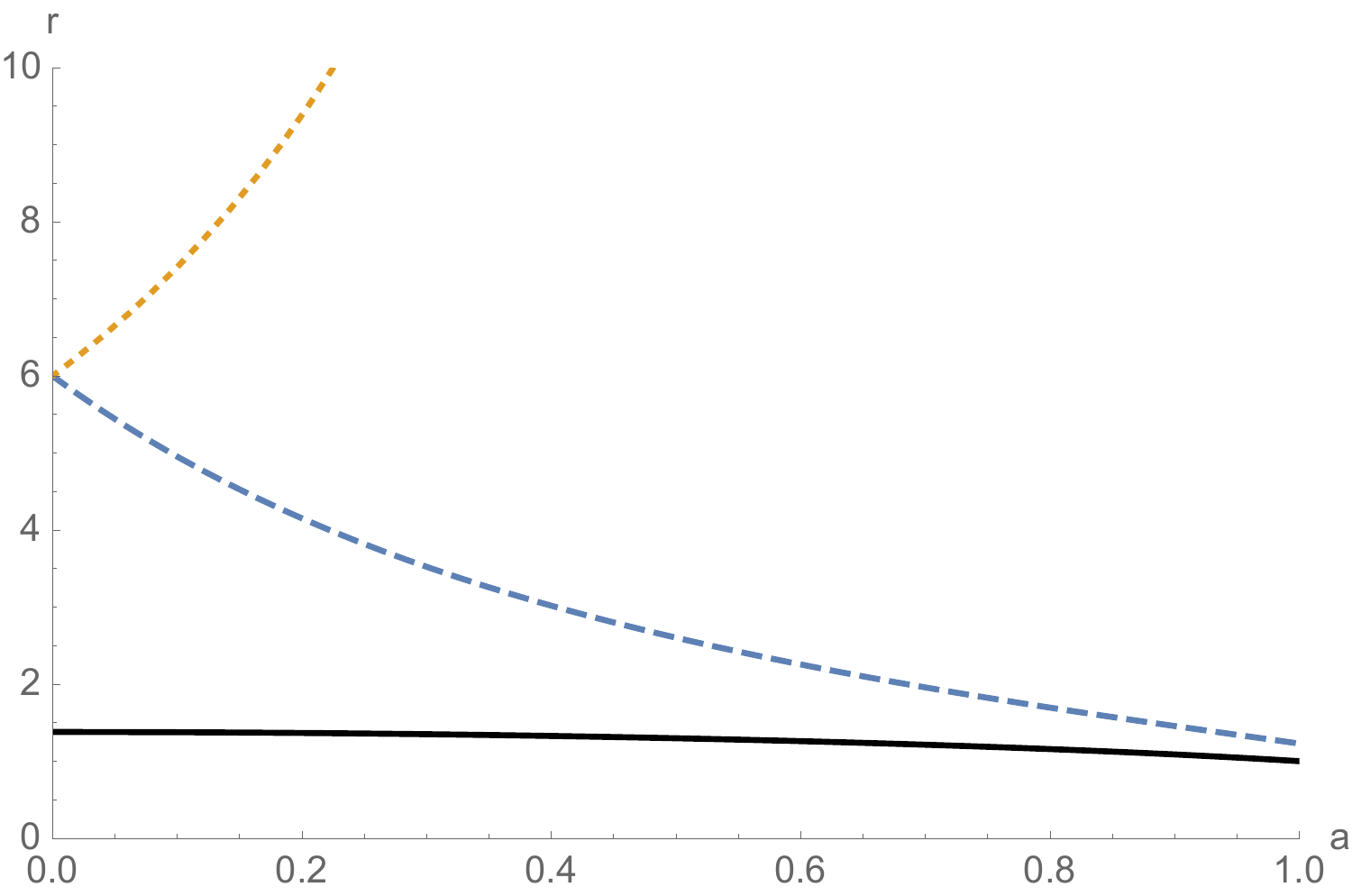}
\end{center}
\caption{The location of the co-rotating (dashed) and counter-rotating (dotted) circular null orbits for the Kerr-AdS black hole as a function of the rotation parameter $a$. We have set the AdS length $L = 1$ and the mass parameter $m=2$. The solid black line is the location of the horizon as a function of the rotation parameter $a$. }
\label{fig:null-circular-orbits}
  \end{figure}

A massless particle moving in the equatorial plane 
$    \theta = \frac{\pi}{2},$ has $\dot{\theta} = 0$ when the Carter's constant is fixed to be,
\be
     K = (aE-J)^2 \chi^2.
\ee
The effective potential for the radial motion then simplifies to,
\be
 \dot{r}^2  +    V_\textrm{eff}(r) = 0,
 \ee
 with
 \be
V_\textrm{eff}(r) :=  - \frac{((r^2 + a^2) E - a J)^2 \chi^2 - \Delta_r K}{r^4}.
\ee
For circular orbits, we have the conditions: $\dot{r}=0$ and $\ddot{r}=0$, which results in
\begin{align}
& V_\textrm{eff}(r)= 0, \\
 & V_\textrm{eff}'(r)=0.
\end{align}
 On eliminating $J$ from these equations we get,
\be
\chi^2 r^3 - 6 m \left(1 + \frac{a^2}{L^2} \right)r^2 + 9 m^2 r - 4 a^2 m = 0.
\ee
This equation has two physical roots (out of three) corresponding to co-rotating and counter-rotating circular photon orbits. For $L=1$ and for $m=2$, the locations of these orbits are plotted in Figure~\ref{fig:null-circular-orbits}.  At the locations of the circular null orbits, the second derivative of the effective potential is negative, 
\begin{equation}
    V_\textrm{eff}''(r)=- \frac{6Km}{r^5}<0,
\end{equation}
 which shows that the circular null orbits are unstable.

 \subsection{Circular timelike orbits}

For a massive particle moving in the equatorial plane 
$\mu^2=1$,  $\theta = \frac{\pi}{2}, \dot{\theta} = 0$. The Carter’s constant is fixed to be,  cf.~\eqref{thetageo},
\begin{equation}
 K = \chi^2 (aE-J)^2.  \label{Carter_constant_timelike_equatorial}
\end{equation} 
The effective potential for the radial motion now simplifies to,
\be
 \dot{r}^2  +    V_\textrm{eff}(r) = 0,
 \ee
 with
\begin{equation}
    V_\mathrm{eff}(r) =- \frac{1}{r^4}\left[\left((r^2+a^2)E-aJ\right)^2\chi^2-\Delta_r(K+ r^2)\right].
\end{equation} 
 \begin{figure}[t]
\begin{center}
  \includegraphics[width=10cm]{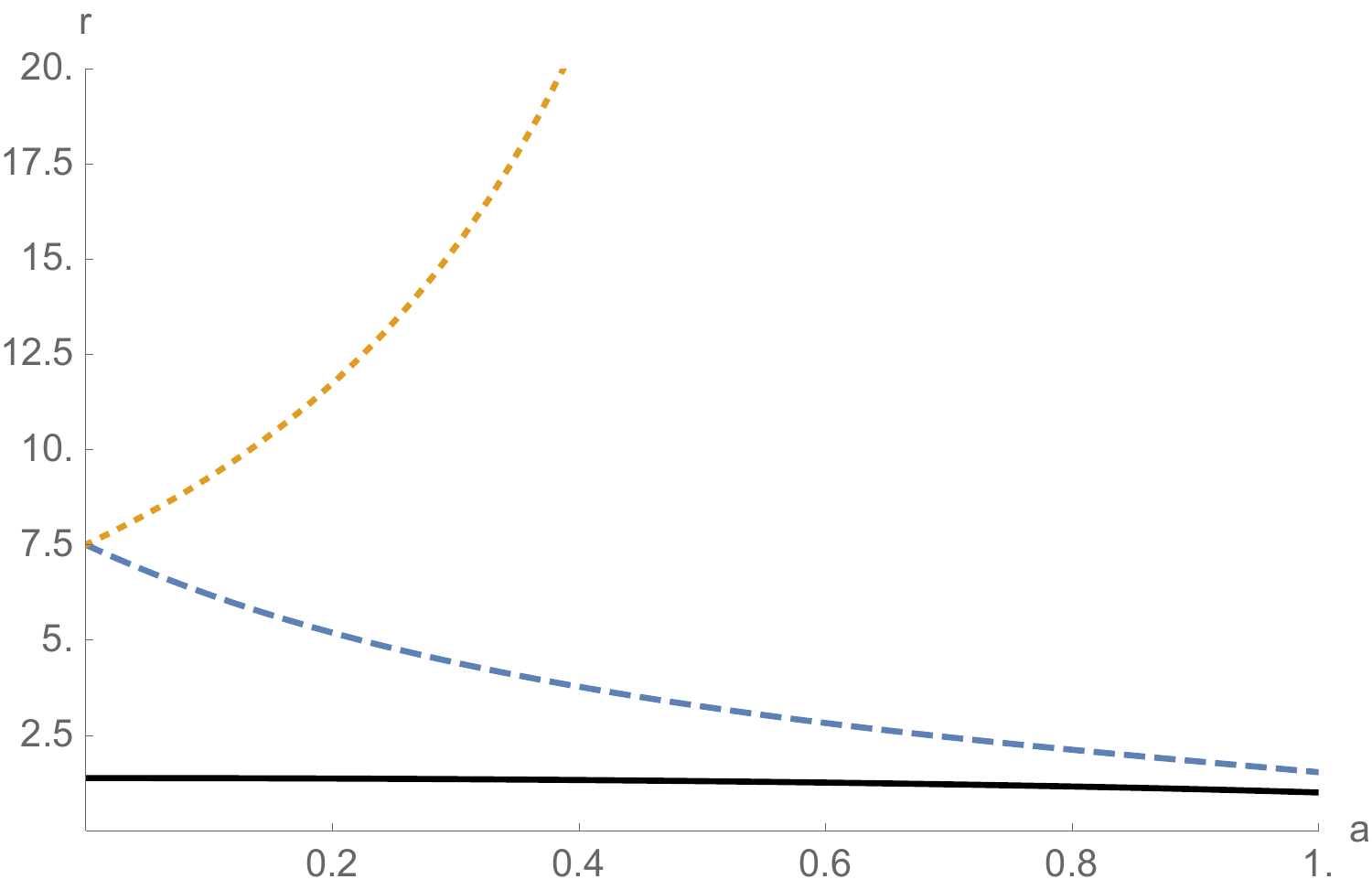}
\end{center}
\caption{The location of the co-rotating (dashed) and counter-rotating (dotted)  timelike circular orbits for the Kerr-AdS black hole as a function of the rotation parameter $a$. We have set the AdS length $L = 1$ and the mass parameter $m=2$ for plotting this graph. The solid black line is the location of the horizon as a function of the rotation parameter $a$. }
\label{fig:timelike-circular-orbits}
  \end{figure} 
For marginally stable circular orbits (see e.g., \cite{Chandrasekhar:1985kt, Hobson:2006se}),
\begin{equation}
    V_\mathrm{eff}(r_\star)=0, \qquad V_\mathrm{eff}'(r_\star)=0, \qquad V_\mathrm{eff}''(r_\star)=0, \label{eqsV}
\end{equation}
where $r_\star$ is the location of these orbits. Equations \eqref{eqsV} can be solved in multiple ways. We approach this problem as follows. First, we eliminate $J$ from the two equations $ V_\mathrm{eff}(r_\star)=0, V_\mathrm{eff}'(r_\star)=0$. This results is a quadratic equation for $K$, solutions to which correspond to the values of $K$ for the co- and counter-rotating orbits.  Next, we consider the combination, 
$3V_\mathrm{eff}'(r)-V_\mathrm{eff}''(r)=0$, which results in a simple equation for $r_\star$,
\begin{equation}
    \frac{3 mLK}{r_\star} = m r_\star L + \frac{4 r_\star^4}{L}.
\end{equation}
This equation can be numerically solved given the values of $K$. This way, we get the innermost stable co- and counter-rotating circular orbits. For AdS length $L=1$ and the mass parameter $m=2$, the locations of these orbits in plotted in Figure~\ref{fig:timelike-circular-orbits}. In our crude model of Grey galaxies, the central black hole has $\Omega L= 1$ and the gas co-rotates around the black hole in stable timelike orbits. The location of the co-rotating innermost stable circular orbit (ISCO) for different values of the rotation parameter is plotted in Figure~\ref{fig:timelike-circular-orbits} as the dashed line. Co-rotating stable timelike orbits exist for all values of $r$ beyond ISCO.

 \section{Parallel-transported orthonormal frame}
 \label{sec:main_results}
 In this section, we show that it is possible to solve the equation of parallel transport 
 \be
 u^\mu \nabla_\mu \lambda^\nu_a =0,
 \ee
applied on an orthonormal tetrad $\lambda^\nu_a$, $a=0,1,2,3$ along an arbitrary timelike or null geodesics with tangent vector $u^\mu$ in the Kerr-AdS spacetime.  Tetrad indices are raised and lowered with $\eta_{ab} = \verb+diag +(-1,1,1,1)$.  Our approach closely follows \cite{Marck1}. There are two main steps involved in the construction. In the first step, we construct a standard locally defined orthonormal frame along the given geodesic with the property that two of its vectors  $\lambda^\nu_0$ and $\lambda^\nu_2$ are parallel transported. In the second step, we obtain the desired parallel transported orthonormal tetrad frame by means of a space rotation by an angle $\Psi$ between the remaining the two vectors.

For the Kerr black hole in flat spacetime, these results find many applications. In AdS, a similar calculation was done for spacelike geodesics in Schwarzschild spacetime in \cite{Dodelson:2020lal} with different motivation.

\subsection{Weyl tensor components in Carter's frame}
We start by writing the Weyl tensor components for the Kerr-AdS metric in the Carter's symmetric tetrad frame $\omega^{(a)}_\mu$. \emph{We display tetrad indices for the Carter's frame in round brackets.} It is defined as,
    \begin{align}
&        \omega^{(0)}=\frac{1}{\chi}\left(\frac{\Delta_r}{\r^{2}}\right)^{\frac{1}{2}}\left(d t-a \sin ^2 \theta d \phi\right), \\
&         \omega^{(1)}=\left(\frac{\r^{2}}{\Delta_r}\right)^{\frac{1}{2}} d r, \\
&        \omega^{(2)}=\left(\frac{\r^{2}}{\Delta_\theta}\right)^{\frac{1}{2}} d \theta,\\
&        \omega^{(3)}=\frac{\sin\theta}{\chi}\left(\frac{\Delta_\theta}{\r^{2}}\right)^{\frac{1}{2}}\left(a d t-\left(r^2+a^2\right) d \phi\right).
    \end{align}
The metric tensor \eqref{metric} then takes the form $ ds^2 = \eta_{(a)(b)}\omega^{(a)}\omega^{(b)}. $
With these definitions, the Weyl tensor $C^{(a)}{ }_{(b)(c)(d)}$ can be presented in the following form,
    \begin{align}
&        \tensor{\Omega}{^{(1)}_{(2)}}=-I_1 \: \omega^{(1)} \wedge \omega^{(2)}+I_2 \: \omega^{(0)} \wedge \omega^{(3)},  \\
&        \tensor{\Omega}{^{(0)}_{(3)}}=-I_1 \:  \omega^{(0)} \wedge \omega^{(3)}-I_2 \: \omega^{(1)} \wedge \omega^{(2)}, \\
&       \tensor{\Omega}{^{(0)}_{(1)}}=-2 I_1 \:  \omega^{(1)} \wedge \omega^{(0)}+2 I_2  \: \omega^{(2)} \wedge \omega^{(3)}, \\
&        \tensor{\Omega}{^{(3)}_{(2)}}=-2 I_1 \: \omega^{(2)} \wedge \omega^{(3)}-2 I_2 \: \omega^{(1)} \wedge \omega^{(0)}, \\
&       \tensor{\Omega}{^{(0)}_{(2)}}=I_1 \: \omega^{(2)} \wedge \omega^{(0)}+I_2 \: \omega^{(1)} \wedge \omega^{(3)}, \\
&        \tensor{\Omega}{^{(3)}_{(1)}}=I_1 \: \omega^{(1)} \wedge \omega^{(3)}-I_2 \: \omega^{(2)} \wedge \omega^{(0)}.
    \end{align}
where, 
\be
\Omega^{(a)}{ }_{(b)}=\frac{1}{2} C^{(a)}{ }_{(b)(c)(d)} \omega^{(c)} \wedge \omega^{(d)},
\ee 
and
\bea
I_1 &=& \frac{m r}{ \r^{6}} \left(r^2-3 a^2 \cos ^2 \theta\right), \\
I_2& =& \frac{m a \cos \theta}{ \r^{6}} \left(3 r^2-a^2 \cos ^2 \theta\right) .
\eea

\subsection{Orthonormal frame for timelike geodesics}

Let the desired parallel-propagated tetrad frame be $\lambda=\left(\lambda_0^\mu,\lambda_1^\mu,\lambda_2^\mu,\lambda_3^\mu \right)$, where $\lambda_0^\mu$ is a timelike vector and $\{\lambda_1^\mu,\lambda_2^\mu,\lambda_3^\mu\}$ are three orthonormal spacelike vectors. These tetrad labels should not be confused with the Carter's frame tetrad indices; Carter's frame indices are in round brackets.

It is natural to choose $\lambda_0^\mu = u^\mu$, where  $u^\mu$ is the unit vector tangent to the geodesic, which is by definition parallel transported. The vector $u^\mu$ is simply $u^\mu = \{ \dot{t}, \dot{r}, \dot{\theta}, \dot{\phi} \} $ 
and is normalised as $u^\mu u_\mu =-1$. Expressions for the coordinate derivatives along the geodesic are given in \eqref{tgeo}--\eqref{thetageo}. We wish to write the components of the vector $\lambda_0^\mu$ in the Carter's tetrad frame,
\be
  \lambda_0^{(a)} = \lambda_0^\mu \omega_\mu^{(a)}.
\ee
A simple calculation gives,
    \begin{align}
        \lambda_0^{(0)}&=\frac{\chi}{(\r^{2}\Delta_r )^{1/2}}\left(E\left(r^2+a^2\right)-a J\right), \label{eq:utetrad0} \\
        \lambda_0^{(1)}&=\left(\frac{\r^{2}}{\Delta_r}\right)^{1/2} \dot{r}, \label{eq:utetrad1} \\
        \lambda_0^{(2)}&=\left(\frac{\r^{2}}{\Delta_\theta}\right)^{1/2}\dot{\theta}, \label{eq:utetrad2} \\
         \lambda_0^{(3)}&=\frac{\chi}{(\r^{2}\Delta_\theta )^{1/2}}\left(a E \sin \theta-\frac{J}{\sin \theta}\right) .\label{eq:utetrad3}
    \end{align} 

It was pointed out by Penrose~\cite{Penrose:1973um} that the vector defined as,
\be
L^\mu = f^\mu{}_\nu \lambda_0^\nu ,
\ee
is also parallel transported along the geodesic and is orthogonal to $\lambda_0^\nu$, where $f_{\mu \nu}$ is the famous Killing-Yano tensor.   For the Kerr-AdS solution, the tetrad components of the Killing-Yano tensor can be expressed as \cite{Frolov:2017kze}:
\begin{equation}
    f_{(a)(b)} \omega^{(a)} \wedge \omega^{(b)}=-r \: \omega^{(2)} \wedge \omega^{(3)}+a \cos \theta \: \omega^{(1)} \wedge \omega^{(0)}. 
\label{KillingYano}
\end{equation} 
The Killing-Yano tensor $f_{\mu\nu} = f_{[\mu\nu]}$ satisfies,
\begin{equation}
    \nabla_\sigma f_{\mu \nu}+\nabla_\nu f_{\mu \sigma}=0 .
\end{equation}
The explicit construct of the vector $L$ gives  $L^{(a)}L_{(a)}=K$, and hence we choose 
\be
\lambda_2^{(a)} = \frac{1}{\sqrt{K}}L^{(a)}.
\ee 
$\lambda_2^{(a)}$ is orthogonal to $\lambda_0^{(a)}$. We have,
    \begin{align}
        \lambda_2^{(0)}&=\left(\frac{\r^{2}}{K \Delta_r}\right)^{\frac{1}{2}} a \cos \theta \: \dot{r}, \label{eq:Ltetrad0} \\
        \lambda_2^{(1)}&=\frac{\chi}{ (K \r^{2} \Delta_r)^{1/2}} a \cos \theta\left(E\left(r^2+a^2\right)-a J\right), \label{eq:Ltetrad1} \\
        \lambda_2^{(2)}&=-\frac{\chi}{(K \r^{2} \Delta_\theta)^{1/2}} r\left(a E \sin \theta-\frac{J}{\sin \theta}\right), \label{eq:Ltetrad2} \\
         \lambda_2^{(3)}&=\left(\frac{\r^{2}}{K \Delta_\theta}\right)^{\frac{1}{2}} r \: \dot{\theta} .\label{eq:Ltetrad3}
    \end{align}

Having presented two orthonormal vectors that are parallel transported, it is now natural to introduce two more vectors $\tilde{\lambda}_1^{(a)}$ and $\tilde{\lambda}_3^{(a)}$ that form a complete orthonormal basis:
\be
\left(\lambda_0^{(a)},\tilde{\lambda}_1^{(a)},\lambda_2^{(a)},\tilde{\lambda}_3^{(a)}\right).
\ee
We introduce,
    \begin{align}
        \tilde{\lambda}_1^{(0)} &= \alpha\left(\frac{\r^{2}}{ K \Delta_r}\right)^{\frac{1}{2}} r \dot{r}, \\
       \tilde{\lambda}_1^{(1)} &= \frac{\alpha\chi}{(K \r^{2} \Delta_r)^{1/2}} r\left\{E\left(r^2+a^2\right)-a J\right\},\\
        \tilde{\lambda}_1^{(2)} &= \frac{\beta\chi}{(K \r^{2} \Delta_\theta)^{1/2}} a \cos \theta\left(a E \sin \theta-\frac{J}{\sin \theta}\right),\\
         \tilde{\lambda}_1^{(3)} &= -\beta\left(\frac{\r^{2}}{K\Delta_\theta}\right)^{\frac{1}{2}} a \cos \theta \dot{\theta},
    \end{align}
and
    \begin{align}
        \tilde{\lambda}_3^{(0)} &= \frac{\alpha\chi}{(\r^{2} \Delta_r)^{1/2}}\left\{E\left(r^2+a^2\right)-a J\right\},\\
       \tilde{\lambda}_3^{(1)} &= \alpha\left(\frac{\r^{2}}{ \Delta_r}\right)^{\frac{1}{2}} \dot{r},\\
        \tilde{\lambda}_3^{(2)} &= \beta \left(\frac{\r^{2}}{ \Delta_\theta}\right)^\frac{1}{2} \dot{\theta},\\
         \tilde{\lambda}_3^{(3)} &= \frac{\beta\chi}{(\r^{2}\Delta_\theta)^{1/2}}\left(a E \sin \theta-\frac{J}{ \sin \theta}\right),
    \end{align}
together with
\bea
\alpha & =& \left(\frac{K-a^2 \cos ^2 \theta}{r^2+K}\right)^{\frac{1}{2}} \\ 
 \beta &=& \left(\frac{r^2+K}{K-a^2 \cos ^2 \theta}\right)^{\frac{1}{2}}.
\eea

A quick calculation shows that these four vectors $\left(\lambda_0^{(a)},\tilde{\lambda}_1^{(a)},\lambda_2^{(a)},\tilde{\lambda}_3^{(a)}\right)$ form an orthonormal basis. In fact, this is how 
$\tilde{\lambda}_1^{(a)}$ and $\tilde{\lambda}_3^{(a)}$ are designed. However, this is not quite the desired tetrad frame. The vectors $\lambda_0^{(a)} $ and $\lambda_2^{(a)}$ are parallel transported but  $\tilde{\lambda}_1^{(a)}$ and $\tilde{\lambda}_3^{(a)}$ are not. This can be remedied by introducing an angle $\Psi$ that rotates $\tilde{\lambda}_1^{(a)}$ and $\tilde{\lambda}_3^{(a)}$ into parallel transported vectors. A (long) calculation shows that,
    \begin{align}
        \lambda_1^{(a)}&=\tilde{\lambda}_1^{(a)} \cos \Psi-\tilde{\lambda}_3^{(a)} \sin \Psi, \\
        \lambda_3^{(a)}&=\tilde{\lambda}_1^{(a)} \sin \Psi+\tilde{\lambda}_3^{(a)} \cos \Psi,
    \end{align}
    are the desired vectors provided,
%
\begin{equation}
    \dot{\Psi}=\chi\frac{K^{\frac{1}{2}}}{\r^{2}}\left\{\frac{E\left(r^2+a^2\right)-a J}{r^2+K}+a \frac{\left(J-a E \sin ^2 \theta\right)}{K-a^2 \cos ^2 \theta}\right\}. \label{PsiEq}
\end{equation}
The rotation angle $\Psi$ in $\left( \tilde{\lambda}_1^{(a)}, \tilde{\lambda}_3^{(a)} \right)$ space changes along the geodesic. Equation \eqref{PsiEq} determines the rotation angle. A careful look at the geodesic equations \eqref{rgeo} and \eqref{thetageo} reveals that $\Psi (r(\tau), \theta(\tau))$ can be taken to be of the form $\Psi = F(r)+ G(\theta)$ with
\be
\dot{\Psi} = (\partial_r \Psi) \: \dot{r} +  (\partial_\theta \Psi) \: \dot{\theta} =  (\partial_r F) \: \dot{r} +  (\partial_\theta G) \: \dot{\theta}.
\ee
Substituting $\dot{r}$ and $\dot{\theta}$ from equations \eqref{rgeo} and \eqref{thetageo}, we find 
expressions for $(\partial_r F) $ and $(\partial_\theta G) $. Those expressions can be readily integrated to give,
\be
        F(r)=\pm \chi K^{\frac{1}{2}} \int^r \frac{E\left(r^2+a^2\right)-a J}{r^2+K} \frac{d r}{\left(\left(E\left(r^2+a^2\right)-a J\right)^2\chi^2-\Delta_r\left( r^2+K\right)\right)^{\frac{1}{2}}},
\label{Fr}
\ee
and
\be        G(\theta)=\pm \chi a K^{\frac{1}{2}} \int^\theta \frac{J-a E \sin ^2 \theta}{\left(K-a^2 \cos ^2 \theta\right)} \frac{d \theta}{\left(\Delta_\theta \left(K-a^2 \cos ^2 \theta \right)-\chi^2\left(a E \sin \theta- \frac{J}{\sin \theta}\right)^2\right)^{\frac{1}{2}}}.
\label{Gtheta}
 \ee
The signs in equations \eqref{Fr} and \eqref{Gtheta} are determined by the signs of $\dot{r}$ and $\dot{\theta}$ while taking the square-roots in \eqref{rgeo} and \eqref{thetageo}, respectively. 

\subsection{Tidal tensor}
Curvature causes geodesics deviation, i.e., neighbouring test-particles accelerate relative to one another. This effect can be conveniently  quantified by the so-called tidal tensor. To define an appropriate notion of the tidal tensor, let us consider an extended body. Let the deviation vector between two points on the extended  test be $\epsilon^\mu$. The geodesic deviation equation tells us
\be
\frac{D^2 \epsilon^\mu}{d\tau^2} := u^\rho \nabla_\rho (u^\sigma \nabla_\sigma \epsilon^\mu)=  - R^\mu{}_{\nu \rho \sigma} u^\nu  \epsilon^\rho u^\sigma,
\ee
where $\epsilon^\mu$ is the deviation vector and $u^\mu$ is the tangent to the reference geodesic. 
We consider the test body to be spacelike, so we project $\epsilon^\mu$ along the parallelly propagated spacelike tetrads $\lambda_i^\mu$. It is more convenient to work with the lower indexed vector $\epsilon_\mu$, and define
\be
x_i = \lambda_{i}^\mu \epsilon_\mu =  \lambda_{i}{}^{(a)} \omega_{(a)}{}^\mu  \epsilon_\mu, \qquad i = 1, 2, 3.
\ee
Since  $ \lambda^{(a)}_i $  are parallel transported along the geodesic, the geodesic deviation equation implies
\be
\frac{D^2 x_i}{d\tau^2}  + K_{ij} x^j= 0, \label{geodesic_deviation}
\ee
where
\be
K_{ij} =  R_{\alpha \beta \gamma \delta} \lambda_0^\alpha   \lambda_i^\beta  \lambda_0^\gamma \lambda_j^\delta = R_{(a)(b)(c)(d)} \lambda_0{ }^{(a)} \lambda_i{ }^{(b)} \lambda_0{ }^{(c)} \lambda_j{ }^{(d)}.
\ee
Since, Einstein equations with the cosmological constant fix the Ricci tensor  to be proportional to the metric 
\be
R_{\mu \nu} = \Lambda g_{\mu \nu},
\ee
the Weyl tensor differs from the Riemann tensor by the following term
\be
C_{\mu \nu \rho \sigma} = R_{\mu \nu \rho \sigma} - \Lambda \left( g_{\mu[\rho} g_{\sigma] \nu} - g_{\nu[\rho } g_{\sigma] \mu} \right) +  \frac{4 \Lambda}{3}   g_{\mu[\rho} g_{\sigma] \nu}.
\ee
As a result,  $K_{ij}$ can be written as 
\be
K_{ij} = C_{ij} - \frac{\Lambda}{3} \delta_{ij},
\ee
where $C_{ij}$
is \cite{Marck1}, 
\begin{equation}
    C_{i j}=C_{(a)(b)(c)(d)} \lambda_0{ }^{(a)} \lambda_i{ }^{(b)} \lambda_0{ }^{(c)} \lambda_j{ }^{(d)}.
    \label{TTgen}
\end{equation} 
We call the spatial tensor $C_{ij}$ the tidal tensor. $C_{i j}$ is a symmetric and traceless tensor due to the properties of Weyl tensor.\footnote{We note that the Weyl tensor has the same symmetries as the Riemann tensor and therefore $C_{ij}$ is a spatial tensor. Indices $i,j$ run over $\{1,2,3\}$. A spatial tensor
transforms covariantly under transformations that leave the time coordinate unchanged.} Using the Weyl tensor components, the diagonal and off-diagonal components of tidal tensor are further simplified after a lengthy calculation,
{\small 
\bea
&&    C_{i i}=\left(1-3\left\{\left(\lambda_0{ }^{(0)} \lambda_i{ }^{(1)}-\lambda_0{ }^{(1)} \lambda_i{ }^{(0)}\right)^2-\left(\lambda_0{ }^{(2)} \lambda_i{ }^{(3)}-\lambda_0{ }^{(3)} \lambda_i{ }^{(2)}\right)^2\right\}\right) I_1 \nn \\ && \qquad -6\left(\lambda_0{ }^{(0)} \lambda_i{ }^{(1)}-\lambda_0{ }^{(1)} \lambda_i{ }^{(0)}\right)\left(\lambda_0{ }^{(2)} \lambda_i{ }^{(3)}-\lambda_0{ }^{(3)} \lambda_i{ }^{(2)}\right) I_2,
    \label{TTdiag}
\eea
}
and
{\small 
\bea
  &&  C_{i j} =  3\left\{\left(\lambda_0^{(0)} \lambda_i^{(1)}-\lambda_0^{(1)} \lambda_i^{(0)}\right)\left(\lambda_0^{(1)} \lambda_j^{(0)}-\lambda_0^{(0)} \lambda_j^{(1)}\right) +
     \left(\lambda_0^{(2)} \lambda_i^{(3)}-\lambda_0^{(3)} \lambda_i^{(2)}\right)\left(\lambda_0^{(2)} \lambda_j^{(3)}-\lambda_0^{(3)} \lambda_j^{(2)}\right)\right\} I_1 \nn \\
    & & \qquad - 3\left\{\left(\lambda_0^{(0)} \lambda_j^{(1)}-\lambda_0^{(1)} \lambda_j^{(0)}\right)\left(\lambda_0^{(2)} \lambda_i^{(3)}-\lambda_0^{(3)} \lambda_i^{(2)}\right)+\left(\lambda_0^{(0)} \lambda_i^{(1)}-\lambda_0^{(1)} \lambda_i^{(0)}\right)\left(\lambda_0^{(2)} \lambda_j^{(3)}-\lambda_0^{(3)} \lambda_j^{(2)}\right)\right\} I_2.\nn  \\
     \label{TToffdiag}
\eea
}
The components of the tidal tensor are explicitly written as,
    \begin{align}
        C_{11} & = \left\{1-3 \frac{S T}{K \r^{4}}\left(r^2-a^2 \cos ^2 \theta\right) \cos ^2 \Psi\right\} I_1+6 \frac{S T}{K \r^{4}} a r \cos \theta  \cos ^2 \Psi I_2,\\
        C_{22} & = \left\{1+\frac{3}{K \r^{4}}\left(r^2 T^2-a^2 \cos ^2 \theta S^2\right)\right\} I_1-6\frac{S T}{K \r^{4}} a r \cos \theta  I_2,\\
        C_{33} & = \left\{1- \frac{3 S T}{K \r^{4}}\left(r^2-a^2 \cos ^2 \theta\right) \sin ^2 \Psi\right\} I_1 + 6 \frac{S T}{K \r^{4}} a r \cos \theta  \sin ^2 \Psi I_2,
    \end{align}
and
    \begin{align}
        C_{12} & = 3 \frac{(S T)^{\frac{1}{2}}}{K \r^{4}}\left\{-ar\cos\theta(S+T) I_1+\left(a^2 \cos ^2 \theta S-r^2 T\right) I_2\right\}  \cos \Psi,\\
        C_{13} & =  \frac{S T}{K \r^{4}}\left\{\left(a^2 \cos ^2 \theta-r^2\right) I_1+2 a r \cos \theta I_2\right\} 3 \cos \Psi \sin \Psi,\\
        C_{23} & = 3 \frac{(S T)^{\frac{1}{2}}}{K \r^{4}}\left\{-a r \cos \theta(S+T) I_1+\left(a^2 \cos ^2 \theta S-r^2 T\right) I_2\right\}  \sin \Psi,\\
    \end{align}
where,
\bea
   S & =& r^2+K,\\ 
   T & =& K-a^2 \cos ^2 \theta.
\eea

\subsection{Pseudo-orthonormal frame for null geodesics}
Proceeding as in the case of timelike geodesics, we let the desired parallel-propagated frame along null geodesics be $\lambda = (\lambda_0^\mu, \lambda_1^\mu, \lambda_2^\mu, \lambda_3^\mu)$. We again choose $\lambda_0^\mu = u^\mu$, where $u^\mu = \{\dot{t}, \dot{r}, \dot{\theta}, \dot{\phi}\}$ is the tangent vector to the null geodesic, with $u^\mu u_\mu = 0$, and $\lambda_2^{(a)} = \frac{1}{\sqrt K}L^{(a)}$, where $L^\mu = f^\mu{}_\nu \lambda_0^\nu$, and $K = L^{(a)}L_{(a)}$. 
The null geodesic equations of motion are,
\bea
    \dot{t}&=&\frac{\chi^2}{\r^2}\left[(r^2+a^2)\frac{\{(r^2+a^2)E-aJ\}}{\Delta_r}-a\frac{(aE\sin^2\theta-J)}{\Delta_\theta}\right], \\
    \dot{\phi}&=&\frac{\chi^2}{\r^2}\left[\frac{a}{\Delta_r}\{(r^2+a^2)E-aJ\}-a\frac{(aE \sin^2\theta-J)}{\Delta_\theta  \sin^2\theta}\right],
\eea
and
\bea
 \label{eq:nullrdot}
    \r^{4}\dot{r}^2&=&\{(r^2+a^2)E-aJ\}^2\chi^2-K\Delta_r, \\
 \label{eq:nullthetadot}
    \r^{4}\dot{\theta}^2&=&K \Delta_\theta -\chi^2\left(aE\sin\theta-\frac{J}{\sin\theta}\right)^2.
\eea

Inspired by Marck's construction from \cite{Marck2}, we form the pseudo-orthonormal tetrad 
$$
\left(\lambda_0^{(a)}, \tilde{\lambda}_1^{(a)}, \lambda_2^{(a)}, \tilde{\lambda}_3^{(a)}\right),
$$ 
where the components of $\lambda_0$ and $\lambda_2$ in Carter's frame continue to be given by equations \eqref{eq:utetrad0}--\eqref{eq:utetrad3} and \eqref{eq:Ltetrad0}--\eqref{eq:Ltetrad3} respectively, except with $\dot{r}$ and $\dot{\theta}$ coming from $\eqref{eq:nullrdot}$ and $\eqref{eq:nullthetadot}$. In the null case, we choose,
\bea
        \tilde{\lambda}_1^{(0)} &=& \alpha \lambda_2^{(0)} = \chi\left(\frac{\r^2}{K \Delta_r}\right)^{\frac{1}{2}} r \dot{r}, \\
        \tilde{\lambda}_1^{(1)} &=& \alpha \lambda_2^{(1)} =
        \frac{\chi}{(K \r^2 \Delta_r)^{1/2}} r \left\{E\left(r^2+a^2\right)-a J\right\}, \\
        \tilde{\lambda}_1^{(2)} &=& \beta \lambda_2^{(2)} = 
        \frac{\chi}{(K \r^2 \Delta_\theta)^{1/2}} a \cos{\theta} \left(a E \sin \theta-\frac{J}{\sin \theta}\right), \\
        \tilde{\lambda}_1^{(3)} &=& \beta \lambda_2^{(3)} = -\chi \left(\frac{\r^2}{K \Delta_\theta}\right)^{\frac{1}{2}} a \cos{\theta} \dot{\theta},
\eea
where $\alpha = \frac{r}{a \cos{\theta}}$, $\beta = -\frac{a \cos{\theta}}{r}$, and
\bea
        \tilde{\lambda}_3^{(0)} &=& -\frac{\r^2}{2K}\lambda_0^{(0)} = -\frac{\chi}{2K} \left(\frac{\r^2}{\Delta_r}\right)^{1/2}\{E(r^2 + a^2)- aJ\}, \\ 
        \tilde{\lambda}_3^{(1)} &=& -\frac{\r^2}{2K}\lambda_0^{(1)} = -\frac{\r^3}{2 K \Delta_r^{1/2}}\dot{r}, \\
        \tilde{\lambda}_3^{(2)} &=& \frac{\r^2}{2K}\lambda_0^{(2)} = \frac{\r^3}{2 K}\dot{\theta}, \\
        \tilde{\lambda}_3^{(3)} &=& \frac{\r^2}{2K}\lambda_0^{(3)} = \frac{\chi}{2K}\left(\frac{\r^2}{\Delta_\theta}\right)^{1/2}\left(a E \sin{\theta} - \frac{J}{\sin{\theta}}\right).
\eea
The scalar product matrix for the frame $\left(\lambda_0^{(a)}, \tilde{\lambda}_1^{(a)}, \lambda_2^{(a)}, \tilde{\lambda}_3^{(a)}\right)$ is 
\begin{equation}
    S = \begin{pmatrix}
        0 & 0 & 0 & 1 \\
        0 & 1 & 0 & 0 \\
        0 & 0 & 1 & 0 \\
        1 & 0 & 0 & 0
    \end{pmatrix}.
\end{equation}
$\tilde{\lambda}_1^{(a)}$ and $\tilde{\lambda}_3^{(a)}$ are designed to give the above scalar product matrix. However, as in the time like case, this is not quite the desired tetrad frame. The vectors $\lambda_0^{(a)} $ and $\lambda_2^{(a)}$ are parallel transported but  $\tilde{\lambda}_1^{(a)}$ and $\tilde{\lambda}_3^{(a)}$ are not. This can be remedied by introducing a variable $\Psi$ (in the null case it is not really an angle) that rotates $\tilde{\lambda}_1^{(a)}$ and $\tilde{\lambda}_3^{(a)}$ into parallel transported vectors. A (long) calculation shows that,
\begin{equation}
    \lambda_1^{(a)} = \tilde{\lambda}_1^{(a)} - \Psi \lambda_0^{(a)},
\end{equation}
\begin{equation}
    \lambda_3^{(a)} = \tilde{\lambda}_3^{(a)} + \Psi \tilde{\lambda}_1^{(a)} - \frac{1}{2}\Psi^{2}\lambda_0^{(a)},
\end{equation}
with 
\begin{equation}
    \Psi = \pm\chi\int \frac{E r^2 dr }{\sqrt{\chi^2 (E(r^2 + a^2) - a J)^2 - K\Delta_r}} \pm \chi\int \frac{aE\cos^{2}{\theta}\;d\theta}{\sqrt{K\Delta_\theta^2 - \chi^{2}(J/\sin{\theta} - aE\sin\theta)^2}},
\end{equation}
are the desired parallel transported vectors.
This equation is the analog of equations \eqref{Fr} and \eqref{Gtheta} for the null geodesics.
The frame $\lambda = \left(\lambda_0^{(a)}, \lambda_1^{(a)}, \lambda_2^{(a)}, \lambda_3^{(a)} \right)$ continues to be pseudo-orthonormal, as can be seen by its scalar product matrix 
\begin{equation}
    S' = \begin{pmatrix}
        0 & 0 & 0 & 1 \\
        0 & 1 & 0 & 0 \\
        0 & 0 & 1 & 0 \\
        1 & 0 & 0 & 0
    \end{pmatrix}.
\end{equation}

\subsection{Relation to Penrose limit}
The Penrose plane wave limit associates to a null-geodesic a plane wave metric
 in Brinkmann coordinates
\be
ds^2 = 2dudv + A_{ij}(u)x^i x^j du^2 + dx_1^2 + dx_2^2, \qquad i,j = 1,2,
\ee
characterised by the wave profile $A_{ij}(u)$. The matrix $A_{ij}(u)$ has a covariant characterisation as curvature components with respect to the parallel transported frame \cite{Blau:2003dz,Blau:2006ar}
\be
A_{ij}(u) = -R_{(a)(b)(c)(d)}\lambda_0^{(a)}\lambda_i^{(b)}\lambda_0^{(c)}\lambda_j^{(d)}. \label{profile_matrix}
\ee
A calculation gives,
\be
A_{11} = -A_{22} = \frac{3Kmr(r^4 - 10 a^2 r^2 \cos^2{\theta} + 5 a^4 \cos^{4}{\theta})}{\rho^{10}}, \label{profile_matrix_1}
\ee 
and 
\be 
A_{12} = A_{21} = \frac{3Kma\cos{\theta}(5r^4 - 10a^2 r^2 \cos^{2}\theta + a^4 \cos^4 \theta)}{\rho^{10}}, \label{profile_matrix_2}
\ee
where $u$ is an affine parameter along the null geodesics $r(u), \theta(u)$. In general, this plane wave metric is time-dependent and not diagonal. It is  identical to the plane wave limit of the Kerr black hole recently reported in \cite{Fransen:2023eqj}. Curiously enough, all the $L$ dependences drop out and the answer \eqref{profile_matrix_1}--\eqref{profile_matrix_2} for the Kerr-AdS is identical to the Kerr case. 

\section{Tidal tensor for equatorial geodesics}
\label{sec:equatorial_geodesics}

Now we only focus on timelike geodesics in the equatorial plane. The Carter’s constant is accordingly fixed to be eq.~\eqref{Carter_constant_timelike_equatorial}.
In the equatorial plane, expressions for the tetrad components significantly simplify. It is worth writing the simplified expressions again. They take the form,
\begin{align}
    \lambda_0^{(0)} & = \chi\frac{E(r^2 + a^2) - aJ}{r \sqrt{\Delta_{r}}}, &
    \lambda_0^{(1)} & = \frac{r\dot{r}}{\sqrt{\Delta_{r}}},\\
    \lambda_0^{(2)} &= 0, &
    \lambda_0^{(3)} & = \chi\frac{aE - J}{r},
\end{align}
\begin{align}
    \lambda_2^{(0)} &=0 &
        \lambda_2^{(1)} & =0 ,\\
    \lambda_2^{(2)} & = - \chi \frac{aE - J}{\sqrt{K}}, &
    \lambda_2^{(3)} &=0, 
\end{align}
together with 
\begin{align}
    \tilde{\lambda}_1^{(0)} & = \frac{r^{2}\dot{r}}{\sqrt{\Delta_{r}(r^2 + K)}}, &
    \tilde{\lambda}_1^{(1)} & = \chi\frac{E(r^2 + a^2) - aJ}{\sqrt{\Delta_{r}(r^2 + K)}}, \\
 \tilde{\lambda}_1^{(2)} & = 0,  & \tilde{\lambda}_1^{(3)} & = 0,
\end{align}
and
\begin{align}
    \tilde{\lambda}_3^{(0)} & = \chi\frac{\sqrt{K}(E(r^2 + a^2) - a J)}{r\sqrt{\Delta_{r}(r^2 + K)}}, &
    \tilde{\lambda}_3^{(1)} & = \frac{\sqrt{K}r\dot{r}}{\sqrt{\Delta_{r}(r^2 + K)}},\\
    \tilde{\lambda}_3^{(2)} & = 0, &
       \tilde{\lambda}_3^{(3)} & = \chi\frac{\sqrt{(r^2 + K)}(aE - J)}{r\sqrt{K}}.
\end{align}
Note that, $\lambda_2$ reduces to be the unit vector orthogonal to the equatorial plane. 
The derivative of $\Psi$ along the geodesic takes the form 
\begin{equation}
    \dot{\Psi} = \frac{\chi( E K + a J - a^2 E)}{\sqrt{K} (r^2 + K)}.
\end{equation}
The diagonal components of the tidal tensor in the equatorial plane simplify to 
\bea
    C_{11} &=& \left(1 - 3\frac{(r^2 + K)}{r^2}\cos^{2}{\Psi}\right)\frac{m}{r^3}, \label{c11}\\
    C_{22} &=& \left(1 + 3\frac{K}{r^2}\right)\frac{m}{r^3}, \label{c22} \\
    C_{33} &=& \left(1 - 3\frac{(r^2 + K)}{r^2}\sin^{2}\Psi\right)\frac{m}{r^3}. \label{c33}
\eea
Of the remaining off-diagonal components, only $C_{13}$ is nonzero,
\be
    C_{13} = -3\frac{(r^2 + K)}{r^5}m\cos{\Psi}\sin{\Psi}. \label{c13}
\ee
These expressions are formally identical to the Kerr case. The dependence on the cosmological constant is only through the Carter's constant $K$  \eqref{Carter_constant_timelike_equatorial}.

\section{Tidal forces and  Grey Galaxies}
\label{sec:grey_galaxies}
The Killing vectors of the Kerr-AdS metric in the non-rotating  coordinate frame $(T, r, \theta, \Phi)$ at infinity are, 
\begin{align}
k^\mu &= \frac{\partial}{\partial T}, & l^\mu &= \frac{\partial}{\partial \Phi}.
\end{align}
The Killing energy and angular momentum  are correctly defined in the non-rotating frame for geodesic motion as 
\begin{align}
E_{T} &= -k^\mu u_\mu,&
J_{\Phi} &= l^\mu u_\mu,
\end{align}
where $u^\mu$ is the tangent vector to the geodesic. The metric components in the $(T, \Phi)$ part of the metric change to,
\bea
g_{TT} &=& g_{tt} \chi^2 - \frac{2a}{L^2} g_{t\phi} \chi  + \frac{a^2}{L^4}g_{\phi \phi}, \\
g_{T\Phi} &=& g_{t \phi} \chi - \frac{a}{L^2} g_{\phi\phi}, \\
g_{\Phi\Phi} &=& g_{\phi \phi},
\eea
as a result, we can express $E_{T}$ and $J_{\Phi}$ in terms of $E$ and $J$ as 
\bea
E_{T} &=& - g_{T\mu} u^\mu =  - g_{TT} \dot{T} - g_{T\Phi} \dot{\Phi} = E \chi + \frac{a}{L^2}J, \\
J_\Phi &=& g_{\Phi\mu} u^\mu =  g_{\Phi T} \dot{T}+  g_{\Phi\Phi} \dot{\Phi}  = J.
\eea
In these variables, the value of the Carter's constant for timelike geodesics on the equatorial plane takes the form
\be
K = (a E_T - J_\Phi)^2.
\ee

From equations \eqref{c11}--\eqref{c33} we see that in the large $r$ expansion the leading order tidal force goes as
\be
C \sim \frac{m}{r^3}.
\ee
\emph{We wish to emphasise that the mass parameter $m$ enters this formula and not the ADM mass of the black hole.  In the limit $a \to L$ for fixed $m$, the black hole becomes very massive (the ADM mass diverges), but the tidal forces do not scale with the ADM mass.}  In this sense, the tidal forces remain small even for a large black hole.

The bulk gas in the Grey Galaxies can be thought of  as a large but uniformly flat pancake that lives in the equatorial plane. The gas that contributes significantly is localised at parametrically large radius of the order
\be
\frac{r}{L} \sim \frac{1}{G_N^\frac{1}{4}}.
\ee
The large radius ensures that the energy density of the gas is
parametrically small. This justifies our modelling of the quantum gas as a collection of timelike particles. 
For timelike geodesics relevant for Grey Galaxies, the magnitude of tidal forces scale as
\be
C \sim G_N^{\frac{3}{4}}.
\ee
As mentioned in the introduction, in the small $G_N$ expansion scheme of \cite{Kim:2023sig} the effects are small. Our analysis provides a way to quantify them.

\section{Conclusions}

\label{sec:conclusions}

One of the most interesting open questions in the black hole perturbation physics is what is the end point of the Kerr-AdS superradiant instability. In a recent paper \cite{Kim:2023sig}, it is conjectured that the endpoint is a Grey Galaxy: a new class of coarse-grained solutions to Einstein’s equations made up of a black hole with critical angular velocity $\Omega L = 1$ at the centre of AdS, surrounded by a large flat disk of thermal bulk gas that revolves around the black hole.

In this paper, we modelled the gas as a collection of particles moving on timelike geodesics. This model -- admittedly crude -- allows us to estimate gravitational forces acting on the gas due to the central black hole. This effect has been neglected in the construction so far, for good reasons. However, as emphasised in the introduction, a complete solution to the problem must also take  these effects into account. Our computations are a small step in that direction.

We embarked on this computation with the additional motivation that computing tidal forces in Kerr-AdS is also of some intrinsic interest. The special separability properties  of Kerr-AdS spacetime allowed us to obtain an explicit solution to the problem of constructing a parallel-transported orthonormal frame along an arbitrary time-like or null geodesic. This enabled the computation of tidal forces on particles moving along these geodesics.

We hope that our computations will be of value in the future discussions of Grey Galaxies.

\subsection*{Acknowledgements}
The work of PRP was supported by CMI undergraduate summer internship program.   We thank Vineeth Krishna, Madhu Mishra, and Chintan Patel for discussions. We  especially thank
Shiraz Minwalla  for his patient explanations on various aspects of Grey Galaxies, and for encouraging us to write up this work. We also thank Jorge Rocha for carefully reading through an earlier draft. 

\appendix

\section{A simple derivation for the ISCO radius for Kerr}
\label{sec:Kerr_ISCO}

Timelike geodesics for the Kerr black hole can be studied by setting $\Lambda =0$ and $\mu^2=1$ in equations \eqref{tgeo}--\eqref{thetageo}. 
 For equatorial geodesics \be \theta = \pi/2, \dot{\theta} = 0, K = (aE-J)^2. \ee The effective potential as a function of radial coordinate takes the form,
\be
 \dot{r}^2  +    V_\textrm{eff}(r) = 0,
 \ee
 where,
\begin{equation}
    V_\mathrm{eff}(r) :=- \frac{1}{r^4}\left[\left((r^2+a^2)E-aJ\right)^2-\Delta_r(K+ r^2)\right].
\end{equation} 
For a particle to remain in circular orbit at radius $r=r_\mathrm{c}$, we require $\dot{r}=0$ and $\ddot{r}=0$. This is equivalent to
\bea
    \label{kerrtimecond1}
 &&   V_\mathrm{eff}(r_\mathrm{c})=0,\\
  &&  V_\mathrm{eff}'(r_\mathrm{c})=0.
    \label{kerrtimecond2}
\eea
Eliminating $J$ from the above two equations, we get a quadratic equation for $K$,
\begin{equation}
    a^4 r_\mathrm{c}^3 -2 a^2 (2 K^2 m + m r_\mathrm{c}^4 + K r_\mathrm{c}^2 (m + r_\mathrm{c})) + r (3 K m - K r_\mathrm{c} + m r_\mathrm{c}^2)^2 = 0
\end{equation}
Solving for $K$, we get
\bea
    K &=& \frac{r_\mathrm{c}^{3/2}(a + \sqrt{m r_\mathrm{c}})^2}{-2 a \sqrt{m} + \sqrt{r_\mathrm{c}} (-3 m + r_\mathrm{c})},  \\
    K &=&  \frac{r_\mathrm{c}^{3/2}(a - \sqrt{m r_\mathrm{c}})^2}{2 a \sqrt{m} + \sqrt{r_\mathrm{c}} (-3 m + r_\mathrm{c})}.
\eea
For marginally stable circular orbits, in addition to \eqref{kerrtimecond1}--\eqref{kerrtimecond2} we also have  $V_\mathrm{eff}''(r_\mathrm{c})=0$ \cite{Chandrasekhar:1985kt, Hobson:2006se}. The equations for $V_\mathrm{eff}'(r_\mathrm{c})=0$  and $V_\mathrm{eff}''(r_\mathrm{c})=0$ are given as respectively:
\bea
 &&   2 J\sqrt{K} + K + \frac{3 K m}{r_\mathrm{c}} + m r_\mathrm{c} -a^2 = 0,\\
    \label{rd2}
  &&    6 J \sqrt{K} + 3 K +  \frac{12 K m}{r_\mathrm{c}} + 2 m r_\mathrm{c} - 3 a^2 = 0.
      \label{rdd2}
\eea
Taking the linear combination of the form $3V_\mathrm{eff}'(r_\mathrm{c})-V_\mathrm{eff}''(r_\mathrm{c})=0 $, results in an equation for $r_\mathrm{c}$,
\be
    \frac{3Km}{r_\mathrm{c}} = m r_\mathrm{c}
\ee
Substituting the two values of $K$ gives the implicit equations for the radius of the innermost stable circular orbits (ISCO):
\be
    r_\mathrm{c}^2 - 6m r_\mathrm{c} - 3a^2 \pm 8a\sqrt{m r_\mathrm{c}} = 0,
\ee
where a positive sign corresponds to the co-rotating orbit and a negative sign to the counter-rotating orbit \cite{Chandrasekhar:1985kt, Hobson:2006se}. In the limit $a = 0$, the two orbits coincide to give $r_\mathrm{ISCO} = 6m$.

\end{document}